\documentstyle[aps]{revtex}
\begin{document}
\begin{title}
{\bf
%Effect of Nonlocality and Edge States Screening on Edge Magnetoplasmons
Random-phase Approximation Treatment Of Edge Magnetoplasmons:
Edge-state Screening And  Nonlocality}
%\end{center}
\end{title}
%\end{center}
\author{O. G. Balev$^{*}$ and P. Vasilopoulos$^\dagger$}
\address{$^{*}$Institute of Physics of Semiconductors, National\\
Academy of Sciences, \\
45 Pr. Nauky, Kiev 252650, Ukraine\\
$\ ^\dagger$Concordia University, Department of Physics, \\
1455 de\\
Maisonneuve Blvd O, Montr\'{e}al, Qu\'{e}bec, Canada, H3G 1M8}
\date{July 6, 1998}
%\title{}
\maketitle

\begin{abstract}
A random-phase approximation (RPA) treatment of edge magnetoplasmons (EMP) 
is presented for strong magnetic
fields, low temperatures, and integer filling factors $\nu$. It is valid
for negligible dissipation and lateral confining potentials
%, flat in the interior of the channel, is assumed
smooth on the scale of the magnetic length $\ell_{0}$ but sufficiently steep
that the Landau level (LL) flattening can be neglected. LL coupling,
screening by edge states, and nonlocal contributions to the current density
are taken into account. In addition to the fundamental mode with typical
dispersion relation $\omega\sim q_x \ln(q_{x})$, fundamental modes with {\it
acoustic} dispersion relation $\omega\sim q_x$ are obtained for $\nu>2$. For
$\nu=1,2$ a {\bf dipole} mode exists, with dispersion relation $\omega\sim
q_x^3$, that is directly related to nonlocal responses.

\ \newline
PACS\ \ 73.20.Dx, 73.40.Hm
\end{abstract}

\section{INTRODUCTION.}

Edge magnetoplasmons (EMP's) have atracted considerable interest in the past
years \cite{1}-\cite{11} and Ref. \cite{1} contains a discussion of older
studies. In \cite{11} we have treated EMPs, $\propto A(\omega,q_{x},y) \exp[
-i(\omega t-q_{x} x)]$, for $\nu=1,2$ and very low temperatures when the
unperturbed density profile drops sharply at the edge on a length of the
order of the magnetic length $\ell_{0}$. Such a profile is valid for $
k_{B}T\ll \hbar v_{g}/\ell_{0}$, where $v_{g}$ is the group velocity of the
edge states. Then for $\nu=1,2$ the unperturbed electron density $n_{0}(y)$,
normalized to the bulk value $n_{0}$, is calculated as $n_{0}(y)/n_{0}=\{1+
\Phi[(y_{re}-y)/\ell_{0}]\}/2$, where $y_{re}$ is the coordinate of the
right edge and $\Phi(y)$ the probability integral.

The main results of Ref. \cite{11} were the existence of many {\it symmetric}
and {\it antisymmetric} modes with respect to the edge, some of which can
propagate nearly undamped even when the dissipation is strong and were,
therefore, termed {\it edge helicons}. In this {\it quasi-microscopic}
treatment, however, the contributions to the current density
$j_{\mu}(\omega,q_{x},y)$ were obtained from microscopic expressions valid
when the components of the electric field ${\bf E}$, associated with the
wave, are smooth on the $\ell_{0}$ scale. This is not well justified.
In Ref. \cite{11} we neglected possible nonlocal effects and approximated the
contributions to $j_{\mu}(\omega,q_{x},y)$ with those obtained when
$E_{y}(y)$ is smooth on the scale of $\ell_{0}$.
The model \cite{11} also neglects
the screening of the two-dimensional electron gas (2DEG). Here, using a RPA
framework we study the effect of nonlocality and of edge-states screening on
EMPs for filling factors $6\geq\nu\geq 1$. For even filling factors $\nu$ we
neglect the spin splitting. As we will show, the existence of additional
modes for $\nu=1(2)$, predicted in Ref. \cite{11}, is confirmed by the
present {\it fully microscopic} treatment and the results for the
fundamental mode remain valid for weak dissipation despite the neglect of
screening and nonlocality in \cite{11}.

In Sec. II we present the channel-edge characteristics and derive an
integral equation for the inhomogeneous wave charge density within a RPA
framework. In Sec. III we obtain the fundamental and dipole EMP's for
$\nu=1,2$ and in Sec. IV we consider fundamental EMPs for $\nu=4$
and $\nu=6$. Finally, in Sec. V we discuss briefly the results and
make concluding remarks.

\section{BASIC RELATIONS}

\subsection{Channel-edge characteristics}

We consider a zero-thickness 2DEG of width $W$ and of length $L_x=L$ in the
presence of a strong magnetic field $B$ along the $z$ axis. We take the
confining potential flat in the interior of the 2DEG ($V_{y}=0$) and parabolic
at its edges, $V_{y} =m^*\Omega^2 (y-y_r)^2/2$, $y\geq y_r$. $V_{y}$ is
assumed smooth on the scale of $\ell_{0}=(\hbar/m^{*} \omega_{c})^{1/2}$
such that $\Omega \ll \omega_{c}$, where $\omega_{c}=|e|B/m^* $ is the
cyclotron frequency ($e<0$). In the Landau gauge for the vector potential
${\bf A}=(-By, 0, 0)$ the one-electron Hamiltonian $\hat{h}^0$ is given by

\begin{equation}
\hat{h}^0=[(\hat{p}_x +eBy)^2 +\hat{p}_{y}^{2}]/2m^* +V_{y} \;\; , \label{1}
\end{equation}
where $\hat{{\bf p}}$ is the momentum operator. The eigenvalues and
eigenfunctions corresponding to Eq. (\ref{1}) near the right edge of the
channel, with $y_{0} \equiv y_{0}(k_{x})=\ell_{0}^{2} k_{x} \geq y_{r}$, are
well approximated by

\begin{equation}  
E_{\alpha}\equiv E_{n}(k_{x})= (n+1/2)\hbar \omega_{c}+m^*\Omega^2
(y_{0}-y_{r})^2/2 , \\    \label{2}
\end{equation}
and
\begin{equation}  
|\alpha> \equiv \psi_\alpha({\bf r}) =e^{ik_x x}\Psi_n(y-y_{0})/\sqrt{L}, \\
\label{3}
\end{equation}
respectively. Here ${\bf r}=\{x,y\}$, $\alpha\equiv \{n,k_x\}$, $\Psi_n(y)$
is a harmonic oscillator function. 
The energy spectrum (\ref{2}) of the n-th LL leads to the group velocity of
the edge states $v_{gn}=\partial E_{n}(k_{r}+k_{e}^{(n)})/\hbar\partial
k_{x}=\hbar \Omega^{2}k_{e}^{(n)}/m^{*}\omega_{c}^{2}$ with characteristic
wave vector $k_{e}^{(n)}=(\omega_{c}/ \hbar\Omega) \sqrt{2m^{*}\Delta_{Fn}}$,
$\Delta_{Fn}=E_{F}-(n+1/2)\hbar \omega_{c}$, and $E_F$ is the Fermi
energy. The edge of the n-th LL is denoted by $y_{rn}=y_{r}+
\ell_{0}^{2}k_{e}^{(n)}=\ell_{0}^{2}k_{rn}$,
where $k_{rn}=k_{r}+k_{e}^{(n)}$,
and $W=2y_{r0}$. We can also write $v_{gn}=E_{en}/B$, where $E_{en}=\Omega
\sqrt{2m^{*}\Delta_{Fn}}/|e|$ is the electric field associated with the
confining potential $V_{y}$ at $y_{rn}$. We also introduce the wave vector
$k_{r}=y_{r}/\ell_{0}^{2}$.

For definiteness, we take the background dielectric constant $\epsilon$ to
be spatially homogeneous. Assuming $|q_{x}| W \gg 1$, we can consider an EMP
along the right edge of the channel of the form $A(\omega, q_{x}, y) \exp[
-i(\omega t-q_{x} x)]$ totally independent of the left edge. For $\nu$ even
the spin-splitting is neglected.

\subsection{ Wave charge density and electric potential at the channel edge}

As in Refs. \cite{11} and \cite{12}, we assume that without interaction the
one-electron density matrix $\hat{\rho}^{(0)}$ is diagonal, i.e., $<\alpha|
\hat{\rho}^{(0)}|\beta>=f_{\alpha}\delta_{\alpha\beta}$, where
$f_{\alpha}=1/[1+exp((E_{\alpha}-E_F)/k_BT)]$ is the Fermi-Dirac function.

For the application of the RPA we follow the self-consistent field approach
of Ref. \cite{13} and references cited therein. The one-electron Hamiltonian
in the presence of a self-consistent wave potential $V(x,y,t)=V(%
\omega_{0},q_{x},y) \exp[-i(\omega_{0} t-q_{x} x)]+ c.c.$, is $\hat{H}(t)=%
\hat{h}^{0}+V(x,y,t)$. The corresponding equation of motion for the
one-electron density matrix $\hat{\rho}$ reads

\begin{equation}
i\hbar\frac{\partial \hat{\rho}}{\partial t}=[\hat{H}(t),\hat{\rho}]%
-\frac{i\hbar}{\tau}(\hat{\rho}-\hat{\rho}^{(0)}),   \label{5}
\end{equation}
where $[,]$ denotes the commutator. On the right hand side (RHS) of Eq.
(\ref{5}) we have introduced the phenomenological infinitesimal term
$\propto 1/\tau$ ($\tau \rightarrow \infty$) that leads to the correct
rules for contour integration around the singularities, cf. Refs. \cite{13}
and \cite{14}. Notice that $\tau \rightarrow \infty$ corresponds to the
collisionless case while using a finite $\tau$ provides the possibility of
estimating roughly the influence of collisions. Here the most effective
collisions are related to intra-LL and intra-edge transitions,
cf. Refs. \cite{11}-\cite{12}.

We take the Laplace trasform, with respect to the time t, of Eq. (\ref{5}),
write $\hat{R}(\omega)=\int_{0}^{\infty}e^{i\omega t} \hat{\rho} dt$ and
$R_{\alpha \beta}(\omega)=<\alpha|\hat{R}(\omega)|\beta>$, and look for
its solution in a power series in $V$

\begin{equation}
R_{\alpha \beta}(\omega)=\sum_{n=0}^{\infty} R_{\alpha \beta}^{(n)}(\omega),
\label{6}
\end{equation}
where $R_{\alpha \beta}^{(0)}(\omega)=(i f_{\alpha}/\omega) \delta_{\alpha
\beta}$. Because we consider linear EMPs, it is sufficient  to take into
account only the first two terms $n=0$ and $n=1$ on the RHS of
Eq. (\ref{6}).  Then in $V(x,y,t)$ we can consider only the term
$V(\omega_{0},q_{x},y) \exp[-i(\omega_{0} t-q_{x} x)]$, which leads to

\begin{equation}
R_{\alpha \beta}^{(1)}(\omega)=\frac{i(f_{\beta}-f_{\alpha})
<\alpha|V(\omega_{0},q_{x},y) e^{iq_{x} x}|\beta>} {(\omega-\omega_{0})
[E_{\beta}-E_{\alpha}+\hbar \omega+i\hbar/\tau]} .  \label{7}
\end{equation}
Taking the trace of $\hat{\rho}$ with the electron charge density operator,
$e \delta({\bf r}-\hat{{\bf r}})$, gives the wave charge density as

\begin{equation}
\delta \rho(t,x,y) \equiv \rho(t,x,y)=\frac{e}{2\pi} \int_{-\infty+i\eta}^{
\infty+i\eta} d \omega e^{-i\omega t} \sum_{\alpha \beta} R_{\alpha
\beta}^{(1)}(\omega) \psi_{\beta}^{*}({\bf r}) \psi_\alpha({\bf r}) ,
\label{8}
\end{equation}
where $\eta>0$. From Eqs. (\ref{7}) and (\ref{8}) it follows that
$\rho(t,x,y)=\rho(t,q_{x},y) \exp(iq_{x} x)$. Moreover, for $t/\tau \gg 1$,
the contributions related to transitional processes are already negligible.
It follows that
$\rho(t,q_{x},y)=\rho(\omega_{0},q_{x},y) \exp(-i\omega_{0}t)$.

Further, from Poisson's equation the wave electric potential $\phi(t,
q_{x},y)$ induced by $\rho(t,q_{x},y)$ is given as %$^{\cite{11}}$

\begin{equation}
\phi(t, q_{x},y)=\frac{2}{\epsilon}\int_{-\infty}^{\infty}
dy^{\prime}K_{0}(|q_{x}||y-y^{\prime}|)\rho(t, q_{x},y^{\prime}), \label{9}
\end{equation}
where $K_{0}(x)$ is the modified Bessel function; $\phi$ and $\rho$ pertain
to the 2D plane.

\subsection{ Integral equation for EMP's}

For $t/\tau \gg 1$ the relation $\rho(t, q_{x},y^{\prime})= \rho(\omega_{0},
q_{x},y^{\prime})\exp(-i\omega_0 t)$ holds; then from Eq. (\ref{9}) it
follows that $\phi(t, q_{x},y)=\phi(\omega_0, q_{x},y)\exp(-i\omega_0 t)$.
In the absence of an external potential we have $V(\omega_{0},q_{x},y)=e
\phi(\omega_0, q_{x},y)$. As a result,  from Eqs. (\ref{7})-(\ref{9}), for
$t/\tau \gg 1$ , we obtain the integral equation for $\rho(\omega, q_{x},y)$
as

\begin{eqnarray}
&&\rho(\omega, q_{x},y)=\frac{2e^{2}}{\epsilon L} \sum_{n_{\alpha},n_{
\beta}=0}^{\infty} \sum_{k_{x\alpha}} \frac{f_{\beta}-f_{\alpha}}{
E_{\beta}-E_{\alpha}+\hbar \omega+ i \hbar/\tau} \
\Pi_{n_{\alpha}n_{\beta}}( y, k_{x\alpha}, k_{x\beta})  \nonumber \\
* \   \nonumber \\
&&\times\int_{-\infty}^{\infty} d\tilde{y} \int_{-\infty}^{\infty}
dy^{\prime}\ \Pi_{n_{\alpha}n_{\beta}}( \tilde{y}, k_{x\alpha}, k_{x\beta})
K_{0}(|q_{x}||\tilde{y}-y^{\prime}|)\ \rho(\omega, q_{x},y^{\prime}) ,
\label{10}
\end{eqnarray}
where $\Pi_{n_{\alpha}n_{\beta}}( y, k_{x\alpha}, k_{x\beta})=
\Psi_{n_{\alpha}}(y-y_0(k_{x\alpha})) \Psi_{n_{\beta}}(y-y_0(k_{x\beta}))$,
$k_{x\beta}=k_{x\alpha}-q_{x}$.
We dropped the subscript $0$ from $\omega_{0}$.

For flat LL's, i.e., for $\Omega \rightarrow 0$ and fixed width of the 2DEG,
we first apply the Fourier transformation along $y$ to Eq. (\ref{10}) and
then carry out the sum over $k_{x\alpha}$ as well as the integral over
$y^{\prime}$. This leads to $\rho(\omega, q_{x},q_{y}) \epsilon_{l}(\omega,
q_{x},q_{y})/\epsilon=0$, where $\epsilon_{l}(\omega,
q_{x},q_{y})=\epsilon_{l}(\omega, {\bf q})  \equiv \epsilon_{l}(\omega, q)$
is the RPA longitudinal dielectric function (${\bf q}=\{q_{x},q_{y}\}$).
That is, for $\rho(\omega, q_{x},q_{y}) \neq 0$ we have

\begin{equation}
\frac{\epsilon_{l}(\omega, {\bf q})}{\epsilon}=1- \frac{2 \pi e^{2}}{%
\epsilon q L^{2}} \sum_{\alpha,\beta} \frac{f_{\beta}-f_{\alpha}}{%
E_{\beta}-E_{\alpha}+\hbar \omega+ i \hbar/\tau} |<\alpha|e^{i\vec q \cdot
\vec r}|\beta>|^{2}=0 .  \label{12}
\end{equation}
This equation gives the RPA dispersion relation for "bulk" longitudinal wave
excitations of a 2DEG in a strong magnetic field. It also shows that the
"bulk" excitations should have $\omega>\omega_{c}$. For definitness, we take
$\omega>0$.

The integral equation (\ref{10}) can be considered as a generalization of
Eq. (10) of Ref. \cite{11} since it takes into account nonlocal
contributions to the current density $\propto \int dy^{\prime}\sigma_{\mu
\gamma}(y,y^{\prime}) E_{\gamma}(y^{\prime})$ and the screening by the edge
and bulk states of the 2DEG. As shown in Ref. \cite{16}, the screening by
edge states can be strong. We will obtain only solutions of Eq. (\ref{10})
that are low in frequency, i.e., solutions for which $\omega/\omega_{c} \ll 1
$. The appearence of any wave branch with $\omega < \omega_{c}$ is
principally related to the existence of the edge of the 2DEG. Hence, any
wave with $\omega < \omega_{c}$ should be localized at the edge and can be
called edge magnetoplasmon.

We consider very low temperatures $T$ satisfying $\hbar v_{gn}\gg \ell_{0}
k_{B} T$. Further, we will assume the long-wavelength limit $q_{x} \ell_{0}
\ll 1$, which is well satisfied, e.g., for fundamental EMP \cite{11} in the
low-frequency region. Then if we compare the terms $\propto f_{\beta^{*}}$,
for given $n_{\beta^{*}}$, on the RHS of Eq. (\ref{10}), we obtain that the
contribution to the sum over $n_{\alpha}$ with $n_{\alpha}=n_{\beta^{*}}$ is
much larger than any other term of this sum or the sum of all terms with
$n_{\alpha} \neq n_{\beta^{*}}$. The small parameter is $|\omega-q_{x}v_{g
n_{\beta^{*}}}(k_{x \beta})|/\omega_{c} \ll 1$, where $v_{g
n_{\beta^{*}}}(k_{x \beta})=\hbar^{-1} \partial E_{n_{\beta^{*}}}(k_{x
\beta})/\partial k_{x \beta}$ is the group velocity of any occupied state
$\{ n_{\beta^{*}},k_{x \beta} \}$ of the $n_{\beta^{*}}$ LL. We assume that
each occupied n-th LL has one intersection with the Fermi level at the edge
of the channel and denote the group velocity of its edge states as $v_{gn}
\equiv v_{gn}(k_{rn})$. The small parameter given above implies $q_{x}
v_{g0}/\omega_{c} \ll 1$, as $v_{g0}$ is typically the largest among
$v_{gn}$. Similar results follow from an analysis of the terms $\propto
f_{\alpha^{*}}$ in the sum over $n_{\beta}$ on the RHS of Eq. (\ref{10}).
Hence, for $\omega \ll \omega_{c}$ and $q_{x} v_{g 0} \ll \omega_{c}$ the
terms with $n_{\alpha} \neq n_{\beta}$ can be neglected. This leads to the
integral equation

\begin{eqnarray}
&&\rho(\omega, q_{x},y)=\frac{2e^{2}}{\epsilon L}
\sum_{n_{\alpha}=0}^{\bar{n}}
\sum_{k_{x\alpha}} \frac{f_{n_{\alpha},k_{x\alpha}-q_x}
-f_{n_{\alpha},k_{x\alpha}}} {E_{n_{\alpha},k_{x\alpha}-q_x}
-E_{n_{\alpha},k_{x\alpha}}+ \hbar \omega+i \hbar/\tau} \
\Pi_{n_{\alpha}n_{\alpha}}( y, k_{x\alpha}, k_{x\alpha}-q_x)  \nonumber \\
* \   \nonumber \\
&&\times\int_{-\infty}^{\infty} d\tilde{y} \int_{-\infty}^{\infty}
dy^{\prime}\ \Pi_{n_{\alpha}n_{\alpha}}( \tilde{y}, k_{x\alpha},
k_{x\alpha}-q_x) K_{0}(|q_{x}||\tilde{y}-y^{\prime}|)\ \rho(\omega,
q_{x},y^{\prime}) ,  \label{14}
\end{eqnarray}
where $\bar{n}$ denotes the highest occupied LL. For even $\nu$ the RHS of
Eq. (\ref{14}) should be multiplied by 2, the spin degeneracy factor. We now
study EMPs following from Eq. (\ref{14}).

\section{EMPs FOR $\protect\nu=1$($2$)}

We first consider the case $\nu=1$ and then explain how the pertinent
formulas should be  modified for $\nu=2$. For $\nu=1$ we have $\bar{n}=0$
and Eq. (\ref{14}) takes the form

\begin{eqnarray}
&&\rho(\omega, q_{x},y)=\frac{e^{2}}{\pi \hbar \epsilon} \int_{-\infty}^{%
\infty} dk_{x\alpha} \frac{f_{0,k_{x\alpha}-q_{x}}-f_{0,k_{x\alpha}}} {%
\tilde{\omega}-v_{g0}(k_{x \alpha}) q_{x}} \ \Pi_{00}( y, k_{x\alpha},
k_{x\alpha}-q_{x})  \nonumber \\
* \   \nonumber \\
&&\times\int_{-\infty}^{\infty} d\tilde{y} \int_{-\infty}^{\infty}
dy^{\prime}\ \Pi_{00}( \tilde{y}, k_{x\alpha}, k_{x\alpha}-q_{x}) \
K_{0}(|q_{x}||\tilde{y}-y^{\prime}|)\ \rho(\omega, q_{x},y^{\prime}) ,
\label{15}
\end{eqnarray}
where $\tilde{\omega}=\omega+i/\tau$. As in Ref. \cite{11}, we seek a
solution of Eq. (\ref{15}) in the form

\begin{equation}
\rho_{0}(\omega,q_{x},y)=\Psi_{0}^{2}(\bar{y}) \sum_{n=0}^{\infty}
\rho^{(n)}_{0}(\omega,q_{x}) H_{n}(\bar{y}/\ell_{0}) = \sum_{n=0}^{\infty}
\sqrt{2^{n}n!} \ \rho^{(n)}_{0}(\omega,q_{x})
\Psi_{n}(\bar{y})\Psi_{0}(\bar{y}).  \label{16}
\end{equation}
Here $\bar{y}=y-y_{r0}=y-y_{0}(k_{r0})$ and Eq. (\ref{16}) is the {\it exact}
solution due to the orthogonality of the Hermite polynomials $H_{n}(x)$. The
expansion (\ref{16}) is specific to $\nu=1 (2)$ under the assumed
conditions. This implies that the charge distortion far from the LL edge,
$\agt 10 \ell_{0}$, is much smaller than near this LL edge, $\alt \ell_{0}$.
Physically this assumption can be well justified. We call the terms $n=0, 1,
2,...$, the monopole, dipole, quadrupole, etc. terms in the expansion of
$\rho_{0}(\omega,q_{x},y)$ relative to $y=y_{r0}$. Notice that expansion
(\ref{16}) can also be understood as an expansion in a complete set of
oscillatory wave functions corresponding to all, occupied ($n=0$) and
empty ($n \geq 1$) LLs.

We now multiply Eq. (\ref{15}) by $H_{m}(\bar{y}/\ell_{0})$ and integrate
over $y$ from $y_{r0}-\Delta y_{0}$ to $y_{r0}+\Delta y_{0}$, where $\Delta
y_{0} \agt 3 \ell_{0}$. With the abbreviation $\rho_{0}^{(m)}(\omega,
q_{x})\equiv \rho_{0}^{(m)}$, we obtain

\begin{equation}
\rho_{0}^{(m)}=\frac{e^{2}}{\pi \hbar \epsilon} \int_{-\infty}^{\infty}
dk_{x\alpha} \frac{f_{0,k_{x\alpha}-q_{x}}-f_{0,k_{x\alpha}}}{\tilde{\omega}
-v_{g0}(k_{x \alpha}) q_{x}} \ d_{m0}(q_{x},\delta k_{x\alpha})
\sum_{n=0}^{\infty} \Big(\frac{2^{n}n!}{2^{m}m!}\Big)^{1/2}
b_{n0}^{(0)}(q_{x},|q_{x}|,\delta k_{x\alpha}) \rho_{0}^{(n)},  \label{17}
\end{equation}
where $\delta k_{x\alpha}=k_{x\alpha}-k_{r0}$, $\delta
k_{x\alpha^{^{\prime}}}=\delta k_{x\alpha}-q_{x}$,

\begin{equation}
d_{m0}(q_{x},\delta k_{x\alpha})= \frac{1}{\sqrt{2^{m}m!}}
\int_{-\infty}^{\infty} d\bar{y} \ H_{m}(\bar{y}/\ell_{0}) \ \Pi_{00}( \bar{y%
}, \delta k_{x\alpha}, \delta k_{x\alpha^{^{\prime}}})   \label{18}
\end{equation}
and

\begin{equation}
b_{n0}^{(0)}(q_{x},|q_{x}|,\delta k_{x\alpha})= \int_{-\infty}^{\infty} d
\bar{y}\ \int_{-\infty}^{\infty} d\bar{y}^{\prime}\ \Pi_{00}( \bar{y},
\delta k_{x\alpha}, \delta k_{x\alpha^{^{\prime}}}) K_{0}(|q_{x}||\bar{y}-
\bar{y}^{\prime}|) \Psi_{n}(\bar{y}^{\prime})\Psi_{0}(\bar{y}^{\prime}) \ .
\label{19}
\end{equation}
In both $d_{m0}(q_{x},\delta k_{x\alpha})$ and
$b_{n0}^{(0)}(q_{x},|q_{x}|,\delta k_{x\alpha})$ the first argument
represents the $q_{x}$-dependent term of the argument of the wave function
$\Psi_{0}(\bar{y}-\delta k_{x\alpha^{^{\prime}}})$ and, if replaced by 0, it
means this term can be neglected. We will also use the coefficients $a_{mn}(
q_{x})$ given by \cite{11}

\begin{equation}
a_{mn}( q_{x})=a_{nm} ( q_{x})=\int_{-\infty}^{\infty} dx\
\Psi_{m}(x)\Psi_{0}(x) \int_{-\infty}^{\infty} dx^{\prime}\
K_{0}(|q_{x}||x-x^{\prime}|)\ \Psi_{n}(x^{\prime})\Psi_{0} (x^{\prime}),
\label{20}
\end{equation}
and satisfying $b_{n0}^{(0)}(0,|q_{x}|,0)=a_{n0}(q_{x})$.

\subsection{Fundamental EMP}

Let us approximate the numerator on the RHS of Eq. (\ref{17}) by the first
nonzero term in the expansion over $q_{x}$: $f_{0,k_{x\alpha}-q_x
}-f_{0,k_{x\alpha}} \approx -q_{x} (\partial f_{0,k_{x\alpha}}/\partial
k_{x\alpha})= \delta(k_{x\alpha}-k_{r0}) q_{x}$. Then Eq. (\ref{17}), after
integration over $k_{x\alpha}$, gives

\begin{equation}
\rho_{0}^{(m)}=\frac{e^{2}}{\pi \hbar \epsilon} \ \frac{q_{x}}{\tilde{\omega}%
_{0}} \ d_{m0}(q_{x},0) \sum_{n=0}^{\infty} \Big(\frac{2^{n}n!}{2^{m}m!}\Big)%
^{1/2} \ b_{n0}^{(0)}(q_{x},|q_{x}|, 0)\ \rho_{0}^{(n)},  \label{21}
\end{equation}
where $\tilde{\omega}_{0}=(\tilde{\omega}-v_{g0} q_{x})$. Further, if we
neglect a small term $\propto q_{x} \ell_{0}$
%in $d_{m0}(q_{x},0)$ and $b_{n0}^{(0)}(q_{x},|q_{x}|,0)$
in the argument of the wave function $\Psi_{0}(\bar{y}+\ell_{0}^{2}q_{x})$,
we obtain $d_{m0}(q_{x},0) \approx d_{m0}(0,0)=\delta_{m,0}$ and
$b_{n0}^{(0)}(q_{x},|q_{x}|,0) \approx b_{n0}^{(0)}(0,|q_{x}|,0)=
a_{n0}(q_{x})$. Substituting the former in Eq. (\ref{21}), we obtain
$\rho_{0}^{(m)} \equiv 0$ for $m \geq 1$; thus, only $\rho_{0}^{(0)}$ can be
finite. Then Eq. (\ref{21}) can be rewritten as

\begin{equation}
\rho_{0}^{(0)}=\frac{e^{2}}{\pi \hbar \epsilon} \ \frac{q_{x}}{\tilde{\omega}%
_{0}}\ a_{00}(q_{x})\ \rho_{0}^{(0)}.  \label{22}
\end{equation}
For $\rho_{0}^{(0)} \neq 0$ Eq. (\ref{22}) gives the dispersion relation

\begin{equation}
\omega=q_{x} v_{g0}+\frac{2}{\epsilon} \sigma_{yx}^{0} q_{x} [\ln(\frac{1}{%
|q_{x}|\ell_{0}})+\frac{3}{4}]-\frac{i}{\tau} ,  \label{23}
\end{equation}
where $\sigma_{yx}^{0}=e^{2}/2 \pi \hbar$ for $\nu=1$ and $%
\sigma_{yx}^{0}=e^{2}/\pi \hbar$ for $\nu=2$. We have used the result
$a_{00}(q_{x}) \approx [\ln(1/|q_{x}|\ell_{0})+3/4]$ \cite{11} for
$|q_{x}|\ell_{0} \ll 1$. If we neglect the dissipation, the dispersion
relation (\ref{23}) and the corresponding charge density profile coincide
with the corresponding expressions for the fundamental mode obtained in Ref.
\cite{11}.

Untill now nonlocal effects and screening by the 2DEG have not changed the
results for the fundamental EMP of Ref. \cite{11}. In Ref. \cite{11} it is
shown that the next term that can affect the fundamental mode of the $n=0$
LL is the quadrupole term. Then, if we neglect dissipation $\omega-q_{x}
v_{g0}$ becomes slightly larger, by a factor $[1+0.125/a_{00}^{2}(q_{x})]
\approx 1$. Typically we have $a_{00}(q_{x}) \agt 10$. In addition, the
spatial structure of the fundamental EMP acquires a small quadrupole term
$|\rho_{0}^{(2)}/\rho_{0}^{(0)}| \approx 1/8 a_{00}(q_{x}) \ll 1$ \cite{11}.
It can be shown from Eqs. (\ref{17})-(\ref{19}) that here too the
corrections to the fundamental EMP due to the quadrupole term are very
small; namely, $a_{00}(q_{x})$ on the RHS of Eq. (\ref{23}) should be
changed to $a_{00}(q_{x})+a_{20}(q_{x})q_{x}^{2} \ell_{0}^{2}/2\sqrt{2}
\approx a_{00}(q_{x})-q_{x}^{2} \ell_{0}^{2}/8$. Further, the spatial
structure of the fundamental EMP acquires a small quadrupole term:
$\rho_{0}^{(2)}/\rho_{0}^{(0)}=q_{x}^{2} \ell_{0}^{2}/8 \ll 1$. Thus, for
weak dissipation the results of Ref. \cite{11} for the fundamental EMP are
nearly the same as those of the present microscopic treatment.

\subsection{Dipole EMP}

For a more acurate calculation of the RHS of Eq. (\ref{17}) we must take
into account higher order terms in the expansion over $q_{x}$. As shown
below, when this is done it leads to additional branches. To obtain the
dipole branch correctly we must keep the first three nonzero terms in the
expansion of the numerator of RHS of Eq. (\ref{17}) over $q_{x}$ by writing

\begin{equation}
f_{0,k_{x\alpha}-q_x }-f_{0,k_{x\alpha}} \approx [q_{x}-\frac{q_{x}^{2}}{2!}%
\frac{\partial}{\partial k_{x\alpha}}+ \frac{q_{x}^{3}}{3!}\frac{\partial^{2}%
}{\partial k_{x\alpha}^{2}}] \delta(k_{x\alpha}-k_{r0}).  \label{24}
\end{equation}
Further, we will consider only the first two terms, $n=0$ and $n= 1$ in the
sum of Eq. (\ref{16}) . Then from Eq. (\ref{17})  for $m=0$ and $m=1$ we
obtain the following system of linear equations for $\rho_{0}^{(0)}$ and
$\rho_{0}^{(1)}$

\begin{equation}
\rho_{0}^{(0)}=\frac{e^{2}}{\pi \hbar \epsilon \tilde{\omega}_{0}}
\int_{-\infty}^{\infty} dk_{x\alpha} (f_{0, k_{x\alpha}-q_x
}-f_{0,k_{x\alpha}}) \ d_{00}(q_{x},\delta k_{x\alpha}) \sum_{n=0}^{1} \Big(%
2^{n}n!\Big)^{1/2} \ b_{n0}^{(0)}(q_{x},|q_{x}|,\delta k_{x\alpha})
\rho_{0}^{(n)},  \label{25}
\end{equation}

\begin{eqnarray}
\rho_{0}^{(1)}=\frac{e^{2}}{\pi \hbar \epsilon \tilde{\omega}_{0}}
\int_{-\infty}^{\infty} dk_{x\alpha} (f_{0,k_{x\alpha}-q_x
}-f_{0,k_{x\alpha}}) \ d_{10}(q_{x},\delta k_{x\alpha}) \sum_{n=0}^{1} \Big(%
\frac{2^{n}n!}{2^{1}1!}\Big)^{1/2} \ b_{n0}^{(0)}(q_{x},|q_{x}|,\delta
k_{x\alpha}) \rho_{0}^{(n)}.  \label{26}
\end{eqnarray}
We assumed that, for $T \rightarrow 0$, $v_{g 0}(k_{x \alpha})$ is
independent of $k_{x \alpha}$ in the vicinity of $k_{x \alpha}=k_{r0}$ and
replaced it by $v_{g 0}$. This assumption does not impose any important
restriction on the analysis. Therefore, taking into account Eq. (\ref{24})
we can replace
$1/(\tilde{\omega}- v_{g 0}(k_{x \alpha}) q_{x})$ by
$1/(\tilde{\omega}-q_{x} v_{g 0})$ and place it in front of the
integrals in Eqs. (\ref{25})-(\ref{26}).

In Eq. (\ref{25}) we can assume
$d_{00}(q_{x},\delta k_{x \alpha})=1$ since corrections to it, varying as
$\sim (q_{x} \ell_{0})^{2}$, are not essential. To evaluate the
contribution $\propto \rho_{0}^{(0)}$ on the RHS of Eq. (\ref{25}) it is
sufficient to keep only the first term, $\propto q_{x}$, on the RHS
of Eq. (\ref{24}). This gives

\begin{equation}
\rho_{0}^{(0)}=\frac{e^{2}}{\pi \hbar \epsilon \tilde{\omega}_{0}} \{q_{x}
b_{00}^{(0)}(0,|q_{x}|,0) \rho_{0}^{(0)}+ \sqrt{2} I_{01}^{(1)}(q_{x})
\rho_{0}^{(1)} \} ,  \label{27}
\end{equation}
where

\begin{equation}
I_{01}^{(1)}(q_{x})=\int_{-\infty}^{\infty} dk_{x\alpha} (f_{0,
k_{x\alpha}-q_x }-f_{0,k_{x\alpha}}) b_{10}^{(0)}(q_{x},|q_{x}|,\delta k_{x
\alpha}) ,  \label{28}
\end{equation}
and where $b_{00}^{(0)}(q_{x},|q_{x}|,0)$ has justifiably been approximated
by $b_{00}^{(0)}(0,|q_{x}|,0)$. Before calculating $I_{01}^{(1)}(q_{x})$ we
notice that the approximation for
$d_{00}(q_{x},\delta k_{x \alpha})$ entails the neglect of small corrections
to $I_{01}^{(1)}(q_{x})$ proportional to $(q_{x} \ell_{0})^{k}$ if
$k \geq 4$. Then Eqs. (\ref{24}) and Eq. (\ref{28}) give

\begin{eqnarray}
I_{01}^{(1)}(q_{x})&=& -\frac{1}{\sqrt{2}\ell_{0}} (q_{x} \ell_{0})^{2}
a_{11}(q_{x})+  \nonumber \\
* && +\frac{q_{x}^{2}}{2} \int_{-\infty}^{\infty} dk_{x\alpha}
\delta(k_{x\alpha}-k_{r0}) [1+ \frac{q_{x}}{3}\frac{\partial}{\partial
k_{x\alpha}}] \frac{\partial}{\partial k_{x\alpha}}
b_{10}^{(0)}(q_{x},|q_{x}|,\delta k_{x \alpha}) .  \label{29}
\end{eqnarray}
The corrections $\propto (q_{x} \ell_{0})^{k}$, $k \geq 4$,  we neglected in
the first term on the RHS  are related to the expansion

\begin{equation}
\Psi_{0}(\bar{y}+\ell_{0}^{2}q_{x})=\Psi_{0}(\bar{y})- \frac{1}{\sqrt{2}}
\Psi_{1}(\bar{y}) (q_{x} \ell_{0})+ \frac{1}{4}[\sqrt{2} \Psi_{2}(\bar{y}%
)-\Psi_{0}(\bar{y})] (q_{x} \ell_{0})^{2}+...  \label{30}
\end{equation}
It can be shown that the term $(q_{x}/3)\partial/\partial k_{x\alpha}$ in
the square brackets 
[...] of Eq. (\ref{29}) gives, upon integration over  $k_{x\alpha}$, a
vanishing contribution  since the  remaining double integral 
changes sign upon making the changes $\bar{y}  \rightarrow -\bar{y}$
and $\bar{y}^{\prime}\rightarrow -\bar{y}^{\prime}$. As for the term
corresponding to $1$ in [...], its evaluation gives $q_{x}^{2}\ell_{0}
a_{11}(q_{x})/\sqrt{2}$. Hence, the two finite terms on the RHS of
Eq. (\ref{29}) cancel each other. The final result is $I_{01}^{(1)}(q_{x})=0$
and Eq. (\ref{27}) takes the form of Eq. (\ref{22}).
That is,  in this approximation the fundamental mode is purely {\it monopole}
and is totally independent of {\it dipole} excitations.

We now consider Eq. (\ref{26}). Here we should consider six different
contributions $I_{1n}^{(1)}(q_{x})$, $I_{1n}^{(2)}(q_{x})$, and
$I_{1n}^{(3)}(q_{x})$ for $n=0,1$ related, respectively, to the first
($\propto q_{x}$), second ($\propto q_{x}^{2}$), and third
($\propto q_{x}^{3}$) term on the RHS of Eq. (\ref{24}).
Then Eq. (\ref{26}) can be rewritten as

\begin{equation}
\rho_{0}^{(1)}=\frac{e^{2}}{\pi \hbar \epsilon \tilde{\omega}_{0}}
\sum_{n=0}^{1} \sum_{l=1}^{3} I_{1n}^{(l)}(q_{x}) \rho_{0}^{(n)} ,
\label{31}
\end{equation}
where

\begin{equation}
I_{1n}^{(1)}(q_{x})=\Big(\frac{2^{n}n!}{2^{1}1!}\Big)^{1/2} q_{x}
d_{10}(q_{x},0) b_{n0}^{(0)}(q_{x},|q_{x}|,0) ,  \label{32}
\end{equation}

\begin{equation}
I_{1n}^{(2)}(q_{x})=-\frac{\sqrt{2^{n-1}n! }}{2!} \ q_{x}^{2}
\int_{-\infty}^{\infty} dk_{x\alpha} d_{10}(q_{x},\delta k_{x\alpha})
b_{n0}^{(0)}(q_{x},|q_{x}|,\delta k_{x\alpha}) \frac{\partial}{\partial
k_{x\alpha}}\delta(k_{x\alpha}-k_{r0}) ,  \label{33}
\end{equation}

\begin{equation}
I_{1n}^{(3)}(q_{x})=\frac{\sqrt{2^{n-1}n!}}{3!} \ q_{x}^{3}
\int_{-\infty}^{\infty} dk_{x\alpha} d_{10}(q_{x},\delta k_{x\alpha})
b_{n0}^{(0)}(q_{x},|q_{x}|,\delta k_{x\alpha}) \frac{\partial^{2}}{\partial
k_{x\alpha}^{2}} \delta(k_{x\alpha}-k_{r0}).  \label{34}
\end{equation}

Again we will neglect contributions $\propto (q_{x} \ell_{0})^{m}$ for $m
\geq 4$. In Eq. (\ref{32}) we have $d_{10}(q_{x},0)=-(q_{x} \ell_{0})/\sqrt{2%
}+O((q_{x} \ell_{0})^{3})$ and $%
b_{00}^{(0)}(q_{x},|q_{x}|,0)=a_{00}(q_{x})+O((q_{x} \ell_{0})^{2})$, $%
b_{10}^{(0)}(q_{x},|q_{x}|,0)=-(q_{x} \ell_{0}/\sqrt{2}) \
a_{11}(q_{x})+O((q_{x} \ell_{0})^{3})$. It follows

\begin{equation}
I_{10}^{(1)}(q_{x})=-\frac{(q_{x} \ell_{0})^{2}}{2 \ell_{0}} a_{00}(q_{x}) ,
\;\;\;\;\; I_{11}^{(1)}(q_{x})=\frac{(q_{x} \ell_{0})^{3}}{2 \ell_{0}} \
a_{11}(q_{x}) .  \label{35}
\end{equation}
Evaluating the integral in Eq. (\ref{33}) by parts we obtain

\begin{eqnarray}
I_{1n}^{(2)}(q_{x})&=&\frac{(2^{n-1}n!)^{1/2}}{2!} \ q_{x}^{2}\ \Big\{ \frac{%
d}{d \delta k_{x\alpha}} \ [d_{10}(q_{x},\delta k_{x\alpha})]_{|\delta
k_{x\alpha}=0} \ b_{n0}^{(0)}(q_{x},|q_{x}|,0)+  \nonumber \\
* &&d_{10}(q_{x},0)\ \frac{d}{d \delta k_{x\alpha}} \
[b_{n0}^{(0)}(q_{x},|q_{x}|,\delta k_{x\alpha})]_{|\delta k_{x\alpha}=0} %
\Big\}.  \label{36}
\end{eqnarray}
%%% Takis in above equation \times I just have deleted
This gives the contributions

\begin{equation}
I_{10}^{(2)}(q_{x})=\frac{(q_{x} \ell_{0})^{2}}{2 \ell_{0}} \ a_{00}(q_{x})
, \;\;\;\;\; I_{11}^{(2)}(q_{x})=-\frac{(q_{x} \ell_{0})^{3}}{\ell_{0}} \
a_{11}(q_{x}) .  \label{37}
\end{equation}
Evaluating the integral in Eq. (\ref{34}) by parts we obtain

\begin{equation}
I_{10}^{(3)}(q_{x})=0 , \;\;\;\;\; I_{11}^{(3)}(q_{x})=\frac{2(q_{x}
\ell_{0})^{3}}{3\ell_{0}} \ a_{11}(q_{x}) .  \label{38}
\end{equation}
Using Eqs. (\ref{35}),(\ref{37}) and (\ref{38}) we can rewrite
Eq. (\ref{31}) as

\begin{equation}
\rho_{0}^{(1)}=\frac{e^{2} \ell_{0}^{2} a_{11}(q_{x})} {6\pi \hbar \epsilon
\tilde{\omega}_{0}} q_{x}^{3} \rho_{0}^{(1)} .  \label{39}
\end{equation}
Eq. (\ref{39}) shows that the {\it dipole} branch, similar to that in Ref.
\cite{11}, is not coupled to {\it monopole} excitations of the charge
density. Further, Eq. (\ref{39}), with $a_{11}(q_{x})=1/2$ for
$|q_{x}|\ell_{0} \ll 1$ and precision
$\alt 0.2 \% $ \cite{11}, gives the dispersion relation of the dipole
branch, $\rho_{0}^{(1)}(\omega,q_{x}) \neq 0$,

\begin{equation}
\omega=v_{g 0}q_{x}+\frac{\ell_{0}^{2}}{6 \epsilon} \sigma_{yx}^{0}
q_{x}^{3} -\frac{i}{\tau} ,  \label{40}
\end{equation}
where $\sigma_{yx}^{0}=e^{2}/2 \pi \hbar$ for $\nu=1$ and
$\sigma_{yx}^{0}=e^{2}/\pi \hbar$ for $\nu=2$.

An important difference of Eq. (\ref{40}) from the corresponding result, Eq.
(40), of Ref. \cite{11} for the pure dipole EMP is that here we have
$\omega-v_{g 0}q_{x} \propto q_{x}^{3}$ whereas in Ref. \cite{11} this
becomes $\omega-v_{g 0}q_{x} \propto q_{x}$. That is, the term caused by the
Coulomb interaction has a {\it nonacoustic} behavior if we neglect
dissipation. The calculations demonstrate clearly that the spatial
dispersion effects, although they depend on $q_{x}$, they are essentially
related to the wave structure along the $y$-direction. It is known that
spatial dispersion is directly related to the nonlocality of responses, as
expressed by the dielectric function, etc. \cite{151}. In Eq. (\ref{40})
both nonlocal effects and edge states screening are taken into account.
We point out that in Ref. \cite {17}
a similar dispersion law, $\omega \propto a q_{x}^{3}$, was obtained for
short-range forces, i.e., for an interaction that is essentially not of the
Coulomb type.

\section{Fundamental EMPs for $\protect\nu=4$ and $6$}

For $\nu=2(\bar{n}+1)$ the $n=0$,...,$n=\bar{n}$ LLs have intersections with
the Fermi level at $y_{r0}$,..., $y_{r\bar{n}}$, respectively, and
Eq. (\ref{14}) can be written as

\begin{eqnarray}
\rho(\omega, q_{x},y&)&=\frac{2e^{2}}{\pi \hbar \epsilon} \sum_{n=0}^{\bar{n}%
} \int_{-\infty}^{\infty} d{k_{x\alpha}} \frac{f_{n,k_{x\alpha}-q_x}
-f_{n,k_{x\alpha}}} {\tilde{\omega}-v_{gn}(k_{x \alpha}) q_{x}} \ \Pi_{nn}(
y, k_{x\alpha}, k_{x\alpha}-q_{x})  \nonumber \\
* \   \nonumber \\
&&\times\int_{-\infty}^{\infty} d\tilde{y} \int_{-\infty}^{\infty}
dy^{\prime}\ \Pi_{nn}( \tilde{y}, k_{x\alpha}, k_{x\alpha}-q_{x})
K_{0}(|q_{x}||\tilde{y}-y^{\prime}|)] \rho(\omega, q_{x},y^{\prime}) .
\label{41}
\end{eqnarray}
For confining potentials smooth on the $\ell_{0}$ scale we have $\Delta
y_{m-1,m}=y_{rm-1}-y_{rm} \gg \ell_{0}$, where $m \leq \bar{n}$. For $\nu=4$
there is only one inter-LL length $\Delta y_{0,1}=y_{r0}-y_{r1}$. Further,
making the same  approximations, in the long-wavelength limit $q_{x}\ell_{0}
\ll 1$, as in Sec. IIIA and integrating over $k_{x\alpha}$ in
Eq. (\ref{41}), we obtain

\begin{eqnarray}
\rho(\omega, q_{x},y)&=&\frac{2e^{2}}{\pi \hbar \epsilon} \sum_{n=0}^{\bar{n}%
} \frac{q_{x}}{\tilde{\omega}_{n}} \Psi_{n}^{2}(y-y_{rn})  \nonumber \\
* \   \nonumber \\
&&\times\int_{-\infty}^{\infty} d\tilde{y} \int_{-\infty}^{\infty}
dy^{\prime}[\Psi_{n}^{2}(\tilde{y}-y_{rn}) K_{0}(|q_{x}||\tilde{y}%
-y^{\prime}|)] \rho(\omega, q_{x},y^{\prime}) ,  \label{42}
\end{eqnarray}
where $\tilde{\omega}_{n} \equiv \tilde{\omega}-v_{gn} q_{x}$. It follows
that $\rho(\omega, q_{x},y)$ can be represented by a sum of charges
$\rho_{n}(\omega, q_{x},y)= \rho_{n}^{(0)}(\omega, q_{x}) \Psi_{n}^2(y-y_{rn})
$, localized at the edge of the $n$-th LL, within a region of extent $\sim
\sqrt{2n+1} \ell_{0}$, $n=0,...,\bar{n}$. The result is

\begin{equation}
\rho(\omega, q_{x},y)=\sum_{n=0}^{\bar{n}} \rho_{n}^{(0)}(\omega, q_{x})
\Psi_{n}^{2}(y-y_{rn}) .  \label{43}
\end{equation}
We substitute Eq. (\ref{43}) into Eq. (\ref{42}) and demand that the
coefficients of $\Psi_{n}^{2}(y-y_{rn})$ on both sides of
Eq. (\ref{42})  be equal.
This leads to $\bar{n}+1$ linear homogeneous
equations for $\rho_{n}^{(0)}(\omega, q_{x})$.
%%% Notice that above is valid if between some \rho_{n}(\omega, q_{x},y)
%%%  the overlap is not exponentially small - as it could be for \bar{n}=2

\subsection{$\protect\nu=4$}

Due to $\Delta y_{0,1} \gg \ell_{0}$ we can neglect the exponentially small
overlap between $\rho_{0}(\omega, q_{x},y)$ and $\rho_{1}(\omega, q_{x},y)$.
We will assume $q_{x} \Delta y_{01} \ll 1$.
As shown above, from Eqs. (\ref{42}) and (\ref{43}), with $\bar{n}=1$, we
obtain the system

\begin{equation}
\rho_{0}^{(0)}=\frac{2e^{2}}{\pi\hbar\epsilon} \frac{q_x}{\tilde{\omega}_{0}}
[a_{00}(q_{x})\rho_0^{(0)}+ a_{00}^{11}( q_{x}, \Delta y_{01})\rho_{1}^{(0)}
] ,  \label{44}
\end{equation}

\begin{equation}
\rho_{1}^{(0)}=\frac{2e^{2}}{\pi\hbar\epsilon} \frac{q_x}{\tilde{\omega}_{1}}
[a_{00}^{11}( q_{x}, \Delta y_{01}) \rho_{0}^{(0)}+ a_{11}^{11}( q_{x},
0)\rho_{1}^{(0)} ] ,  \label{45}
\end{equation}
where $\rho_{n}^{(0)}(\omega, q_{x}) \equiv \rho_{n}^{(0)}$ and

\begin{equation}
a_{nn}^{mm}( q_{x}, \Delta y)=
\int_{-\infty}^{\infty}\int_{-\infty}^{\infty} dx\
dx^{\prime}\Psi_{n}^2(x)\Psi_{m}^2(x^{^{\prime}})
K_{0}(|q_{x}||x-x^{\prime}+\Delta y|).  \label{46}
\end{equation}
Here $a_{nn}^{mm}( q_{x}, \Delta y)=a_{mm}^{nn}( q_{x}, \Delta y)$,
$a_{nn}^{mm}( q_{x}, \Delta y)=a_{nn}^{mm}( q_{x}, -\Delta y)$ and
$a_{00}^{00}( q_{x},0)=a_{00}(q_{x})$. For a nontrivial solution of the
system of Eqs. (\ref{44}) and (\ref{45}) the determinant of the coefficients
must vanish. This gives the dispersion relation of the renormalized
fundamental EMP of the $n=0$ LL as

\begin{equation}
\omega=q_{x} v_{01} +\frac{\sigma_{yx}^{0}}{\epsilon} \ q_{x} [2 \ln(\frac{1%
}{|q_{x}|\ell_{0}})-\ln(\frac{\Delta y_{01}}{\ell_{0}})+ \frac{3}{5}]-\frac{i%
}{\tau} ,  \label{47}
\end{equation}
and that of the $n=1$ LL as

\begin{equation}
\omega=q_{x} v_{01} +\frac{\sigma_{yx}^{0}}{\epsilon} \ q_{x} [\ln(\frac{%
\Delta y_{01}}{\ell_{0}})+\frac{2}{5}]-\frac{i}{\tau} ,  \label{48}
\end{equation}
where $\sigma_{yx}^{0}=2e^{2}/\pi \hbar$ and $v_{01} =(v_{g0}+v_{g1})/2$.
All double integrals involved in the coefficients of Eqs. (\ref{44}) and
(\ref{45}) can be evaluated within the approximation $K_{0}(x) \approx
\ln(2/|x|)-\gamma$, where $\gamma$ is the Euler constant. We obtain
$a_{11}^{11}( q_{x}, 0) \approx \ln(1/|q_{x}|\ell_{0})+1/4$ and
$a_{00}^{11}(q_{x}, \Delta y_{01}) \approx
(\ln(2)-\gamma)- \ln(|q_{x}|\Delta y_{01}) \approx \ln(1/|q_{x}\Delta
y_{01}|)+0.1$. Notice that in the absence of dissipation Eqs. (\ref{47}) and
(\ref{48}) coincide with the quasi-microscopic results of Ref. \cite{18}.
>From Eqs. (\ref{44}) and (\ref{45}) it is easy to see that, if the inter-LL
Coulomb coupling is neglected by setting $a_{00}^{11}( q_{x}, \Delta
y_{01})\equiv 0$, the dispersion laws of the {\it decoupled} fundamental
EMPs of the $n=0$ and $n=1$ LLs are given, respectively, by Eq. (\ref{23})
and by

\begin{equation}
\omega=q_{x} v_{g1}+ \frac{2e^{2}}{\pi \hbar \epsilon} q_{x} [\ln(\frac{1}{%
|q_{x}|\ell_{0}})+\frac{1}{4}]-\frac{i}{\tau} .  \label{49}
\end{equation}
Substituting Eq. (\ref{47}) in Eq. (\ref{44}) and Eq. (\ref{48}) in Eq. (\ref
{45}) we obtain, respectively, $\rho_{1}^{(0)}/\rho_{0}^{(0)} \approx 1$
%%%\ln(1/|q_{x}\Delta y_{01}|)+0.1$
and $\rho_{0}^{(0)}/\rho_{1}^{(0)} \approx -1$. This means that in the
former case the wave charges localized at the edges of the $n=0$ LL and $n=1$
LLs are in phase whereas in the latter they are out of phase. Therefore, the
former EMP has an {\it acoustic} spatial structure along the $y$ axis while
the latter EMP has an {\it optical} spatial structure though its dispersion
law is purely acoustic.

\subsection{$\protect\nu=6$}

For $\nu=6$ we have the intersection of the $n=2$ LL with the Fermi level,
at $y_{r2}$, in addition those of the $n=0$ and $n=1$ LLs; $\bar{n}=2$.
Corresponding to Eqs. (\ref{42}) and Eq. (\ref{43}) we now obtain

\begin{equation}
\rho_{0}^{(0)}=\frac{2e^{2}}{\pi\hbar\epsilon} \frac{q_x}{\tilde{\omega}_{0}}
[a_{00}(q_{x})\rho_{0}^{(0)}+ a_{00}^{11}( q_{x}, \Delta
y_{01})\rho_{1}^{(0)}+ a_{00}^{22}( q_{x}, \Delta y_{02})\rho_{2}^{(0)} ] ,
\label{50}
\end{equation}

\begin{equation}
\rho_{1}^{(0)}=\frac{2e^{2}}{\pi\hbar\epsilon} \frac{q_x}{\tilde{\omega}_{1}}
[a_{00}^{11}( q_{x}, \Delta y_{01}) \rho_{0}^{(0)}+ a_{11}^{11}( q_{x}, 0)
\rho_{1}^{(0)}+ a_{11}^{22}( q_{x}, \Delta y_{12}) \rho_{2}^{(0)} ] ,
\label{51}
\end{equation}

\begin{equation}
\rho_{2}^{(0)}=\frac{2e^{2}}{\pi\hbar\epsilon} \frac{q_x}{\tilde{\omega}_{2}}
[a_{00}^{22}( q_{x}, \Delta y_{02}) \rho_{0}^{(0)}+ a_{11}^{22}( q_{x},
\Delta y_{12}) \rho_{1}^{(0)}+ a_{22}^{22}( q_{x}, 0) \rho_{2}^{(0)} ] .
\label{52}
\end{equation}
The vanishing of the determinant of the coefficients leads to the cubic
equation

\begin{equation}
\omega^{` 3}+a_{2}(q_{x}) \omega^{` 2}+ a_{1}(q_{x})
\omega^{`}+a_{0}(q_{x})=0 ,  \label{53}
\end{equation}
where $\omega^{` }=\tilde{\omega}/ (2e^{2}q_{x}/\pi \hbar \epsilon)$. The
expressions for the coefficients $a_{k}(q_{x}), \ k=0,1,2$, are given in the
appendix. We will assume that $q_{x} \Delta y_{ij} \ll 1$, $i \neq j \leq 2$.

It can be shown that all three roots of Eq. (\ref{53}) are real and
different as they correspond to the irreducible case. Assuming $\Delta
y_{02}, \Delta y_{12}, \Delta y_{01} \gg \ell_{0}$,  these roots give the
dispersion law for the renormalized fundamental EMP of the $n=0$ LL as

\begin{equation}
\omega=q_{x} v_{012} + \frac{2\sigma_{yx}^{0}}{3\epsilon} q_{x} \{3 \ln(%
\frac{1}{|q_{x}|\ell_{0}})- \frac{2}{3}[\ln(\frac{\Delta y_{02}}{\ell_{0}})+
\ln(\frac{\Delta y_{01}}{\ell_{0}})+ \ln(\frac{\Delta y_{12}}{\ell_{0}})]+
\frac{8}{15} \}-\frac{i}{\tau} ,  \label{57}
\end{equation}
and that for the renormalized fundamental EMP of the $n=1$ LL, $\omega_{+}$,
and $n=2$ LL, $\omega_{-}$, as

\begin{eqnarray}
\omega_{\pm}=&&q_{x} v_{012} + \frac{2\sigma_{yx}^{0}}{9\epsilon} q_{x}
\{\ln(\frac{\Delta y_{02}}{\ell_{0}})+ \ln(\frac{\Delta y_{01}}{\ell_{0}})+
\ln(\frac{\Delta y_{12}}{\ell_{0}})+\frac{7}{10} \} \pm  \nonumber \\
* \   \nonumber \\
&&\frac{2 \sqrt{2}\sigma_{yx}^{0}}{9\epsilon} q_{x} \{\ln^{2}(\frac{\Delta
y_{02}}{\Delta y_{01}})+ \ln^{2}(\frac{\Delta y_{02}}{\Delta y_{12}})+
\ln^{2}(\frac{\Delta y_{01}}{\Delta y_{12}})+A \}^{1/2}-\frac{i}{\tau} .
\label{58}
\end{eqnarray}
Here $\sigma_{yx}^{0}=3e^{2}/\pi \hbar$,
$v_{012} =(v_{g0}+v_{g1}+v_{g2})/3$, and

\begin{eqnarray}
A=&&(\tilde{v}_{g0}+3/4)[ (\tilde{v}_{g0}- \tilde{v}_{g2}+ 3/4)/2+\ln(\Delta
y_{02}\Delta y_{01}/ {\Delta y_{12}}^{2})]  \nonumber \\
* &&+(\tilde{v}_{g1}+1/4)[(\tilde{v}_{g1}- \tilde{v}_{g0}- 1/2)/2+
\ln(\Delta y_{01}\Delta y_{12}/ {\Delta y_{02}}^{2})]  \nonumber \\
* &&+\tilde{v}_{g2} [ (\tilde{v}_{g2}- \tilde{v}_{g1}-1/4)/2+\ln(\Delta
y_{02}\Delta y_{12}/ {\Delta y_{01}}^{2})] ,  \label{59}
\end{eqnarray}
where $\tilde{v}_{gi}=v_{gi}/(2e^{2}/\pi \hbar \epsilon)$, $i=0,1,2$. Notice
that the fundamental EMPs of the $n=1$ and $n=2$ LLs have purely acoustic
dispersion laws, cf. Eq. (\ref{58}) . Only the fundamental EMP of the $n=0$
LL, Eq. (\ref{57}), behaves in the previously obtained manner $\propto q_{x}
\ln(q_{x})$, cf. Refs. \cite{1}-\cite{10}), for a fundamental EMP. It is
difficult to clearly disentangle the contributions of the $n=2$ LL in
Eq. (\ref{58}). It can be shown though that, if they are neglected by
setting $\tilde{a}_{22}^{22}$, $a_{11}^{22}$, $a_{00}^{22}$, $v_{g2}$
to zero, then we have $\omega^{^{\prime}}_{2}=0$ and Re$\omega_{-}=0$ as
we should in this limit.

The solutions of the cubic equation, Eqs. (\ref{57}) and  (\ref{58}), are
not obtained using the standard expressions,  reproduced in the appendix,
which become very unwieldy for the  present case. Instead, the following
reasoning is used. If $\omega$ has an acoustic character for some branches,
then $\omega^{`}$ in Eq. (\ref{53}) is independent of $q_{x}$ and should
satisfy the quadratic equation obtained by differentiating Eq. (\ref{53})
with respect to $q_{x}$

\begin{equation}
\frac{\partial a_{2}(q_{x})}{\partial q_{x}} \omega^{`2}+ \frac{\partial
a_{1}(q_{x})}{\partial q_{x}} \omega^{`}+ \frac{\partial a_{0}(q_{x})}{%
\partial q_{x}}=0 .  \label{60}
\end{equation}
The two roots $\omega_{2}^{`}$ and $\omega^{`}_{3}$ of Eq. (\ref{60}) are
indeed independent of $q_{x}$ and lead to the solutions $\omega_{-}$ and
$\omega_{+}$, respectively, given by Eq. (\ref{58}). The other root is given
by $\omega^{`}_{1}=-a_{2}(q_{x})-(\omega^{^{\prime}}_{2}+
\omega^{^{\prime}}_{3})$ and results in Eq. (\ref{57}). Of course both
methods give the same solutions $\omega^{^{\prime}}_{j}$, $j=1,2,3$.
Notice that on the RHS of Eqs. (\ref{57}) and (\ref{58}) we  used
{\it approximate} expressions of $a_{ii}^{jj}(q_{x},\Delta y_{ij})$ which ,
assuming $\Delta y_{ij}/\ell_{0} \gg 1$, $i \neq j$, are obtained with the
approximation

\begin{equation}
\int_{-\infty}^{\infty}\int_{-\infty}^{\infty} dx\
dx^{\prime}\Psi_{i}^2(x)\Psi_{j}^2(x^{^{\prime}}) \ln(|x-x^{\prime}+\Delta
y_{ij}|/\ell_{0}) \approx \ln(\Delta y_{ij}/\ell_{0}) .  \label{61}
\end{equation}
Below we will demonstrate that the approximation (\ref{61}) is well
justified even for not-too-large values $\Delta y_{ij}/\ell_{0}$. For that
we will calculate the dispersion laws of the renormalized fundamental EMPs
using the {\it exact} expressions of $a_{ii}^{jj}(q_{x},\Delta y_{ij})$,
i.e., with the double integral on the LHS of Eq. (\ref{61}) evaluated
numerically. In the latter case we will calculate $\omega^{^{\prime}}_{j}$
from the standard expressions (\ref{A1}) -(\ref{A3}) of the appendix.

For a $GaAs$-based 2DEG and negligible dissipation the dispersion laws
(\ref{57}) and (\ref{58}) for the renormalized, by the inter-LL Coulomb
coupling, fundamental EMPs are shown in Fig. 1 by the solid curves. The
parameters are $m^{*} \approx 6.1 \times 10^{-29} g$, $\epsilon \approx 12.5$,
and $\Omega \approx 7.8 \times 10^{11} sec^{-1}$ \cite{19}. For $\nu=6$
and $B=3 $Tesla these parameters lead to $\hbar \omega_{c} \approx 5 meV$,
$\omega_{c}/\Omega \approx 10$. Here $\omega_{*}=2e^{2}/\pi \hbar
\epsilon \ell_{0}$ is a characteristic frequency almost equal, for the
conditions stated,  to $\omega_{c}$. We have also assumed
$\Delta_{F2}=\hbar \omega_{c}/2$. This gives $v_{g0} \approx \sqrt{5}
\Omega \ell_{0}$, $v_{g1} \approx \sqrt{3} \Omega \ell_{0}$, $v_{g2} \approx
\Omega \ell_{0}$, $\Delta y_{02}/\ell_{0} \approx (\omega_{c}/\Omega)
[\sqrt{5}-1]$, $\Delta y_{01}/\ell_{0} \approx (\omega_{c}/\Omega)
[\sqrt{5}-\sqrt{3}]$, and $\Delta y_{12}/\ell_{0} \approx (\omega_{c}/
\Omega)[\sqrt{3}-1]$. The dashed curves show the fundamental EMPs of
the totally decoupled $n=0$ (Eq. (\ref{23}), topmost dashed curve),
$n=1$, and $n=2$ (lowest dashed curve) LLs, i.e., the modes obtained
by neglecting the inter-LL Coulomb coupling. In this case all coefficients
$a_{ii}^{jj}$ on the RHS of Eqs. (\ref{50})-(\ref{52}) vanish for $i \neq j$.
This leads directly to the dispersion laws represented by the dashed curves
in Fig. 1. The dotted curves show the dispersion laws of the renormalized
fundamental EMPs using the exact expressions for $a_{ii}^{jj}(q_{x},\Delta
y_{ij})$, $i \neq j$. It is seen that each dotted curve is very close to the
corresponding solid curve.

For the same parameters and conditions as in Fig. 1  we show, in Fig. 2 ,
$\rho(\omega,q_{x},y)/\rho_{*}$ (solid and dotted curves) and
$\rho(\omega,q_{x},y)/2\rho_{*}$ (dashed curve) as a function of
$Y=(y-y_{r2})/\ell_{0}$ with
$\rho_{*}=\rho_{0}^{(0)}(\omega,q_{x})/\sqrt{\pi} \ell_{0}$. The solid curve
corresponds to the topmost solid curve in Fig. 1, Eq. (\ref{57}), with
$\rho_{1}^{(0)}/\rho_{0}^{(0)} \approx 1.0$ and
$\rho_{2}^{(0)}/\rho_{0}^{(0)} \approx 1$.
As can be seen the renormalized fundamental EMP of the $n=0$ LL has an
{\it acoustic} spatial structure along the $y$ axis. The dashed curve
corresponds to the lowest solid curve in Fig. 1, $\omega_{-}$ of Eq. (\ref
{58}), with $\rho_{1}^{(0)}/\rho_{0}^{(0)} \approx -2.0$ and
$\rho_{2}^{(0)}/\rho_{0}^{(0)} \approx 1$. Hence the renormalized fundamental
EMP of the $n=2$ LL has an {\it optical} spatial structure though its
dispersion law is purely acoustic. The dotted curve corresponds to the
middle solid curve in Fig. 1, $\omega_{+}$ of Eq. (\ref{58}), with
$\rho_{1}^{(0)}/\rho_{0}^{(0)} \approx -0.1$ and $\rho_{2}^{(0)}/
\rho_{0}^{(0)} \approx -1.0$. Thus, the renormalized fundamental EMP
of the $n=1$ LL has an {\it optical} spatial structure though its
dispersion law is purely acoustic. Notice that the dependence of
$\rho_{j}^{(0)}/\rho_{0}^{(0)}$, $j=1,2$, on $q_{x}$ is typically weak.

In Fig. 3 we use the same values for $m^{*}$, $\epsilon$, $\Omega$, and
$\Delta_{F2}=\hbar \omega_{c}/2$, $\nu=6$ as in Fig. 1. However, we take
$B=1.5$Tesla and this leads to $\omega_{c}/\Omega \approx 5$. This gives
$\sqrt{2}$ times larger values for $v_{g0} \approx \sqrt{5} \Omega \ell_{0}$, $v_{g1} \approx
\sqrt{3} \Omega \ell_{0}$, and $v_{g2} \approx \Omega \ell_{0}$, and twice
smaller values for $\Delta y_{02}/\ell_{0} \approx (\omega_{c}/\Omega)[
\sqrt{5}-1]$, $\Delta y_{01}/\ell_{0} \approx (\omega_{c}/\Omega)[\sqrt{5}-
\sqrt{3}]$, and $\Delta y_{12}/\ell_{0} \approx (\omega_{c}/\Omega)[\sqrt{3}
-1]$ as compared with those in Fig. 1. All  curves are marked as in Fig. 1
and show the corresponding  dispersion laws.

In Fig. 4 we plot $\rho(\omega,q_{x},y)/\rho_{*}$ as a function of
$Y=(y-y_{r2})/\ell_{0}$ for the conditions corresponding to the solid curves
in Fig. 3. All curves are marked as in Fig. 2. We have, respectively,
$\rho_{1}^{(0)}/\rho_{0}^{(0)} \approx 1.0$ and
$\rho_{2}^{(0)}/\rho_{0}^{(0)} \approx 1$ (solid  curve),
$\rho_{1}^{(0)}/\rho_{0}^{(0)} \approx -2.1$ and
$\rho_{2}^{(0)}/\rho_{0}^{(0)} \approx 1.2$ (dashed curve), and
$\rho_{1}^{(0)}/\rho_{0}^{(0)} \approx -0.1$ and
$\rho_{2}^{(0)}/\rho_{0}^{(0)} \approx -1.0$ (dotted  curve).
As in Fg. 2, the renormalized fundamental EMP of the $n=0$ LL has an
{\it acoustic} spatial structure along the $y$ axis while those of the
$n=1$ and $n=2$ LL have an {\it optical} spatial structure though their
dispersion laws are purely acoustic. As noted above, the dependence of
$\rho_{j}^{(0)}/\rho_{0}^{(0)}$, $j=1,2$, on $q_{x}$ is typically weak.

\section{discussion and concluding remarks}

We presented a fully {\it microscopic} model of EMPs in a RPA framework
valid for integer $\nu\geq 1$ and confining potentials that are smooth on
the $\ell_{0}$ scale but still sufficiently steep that LL flattening can be
neglected \cite{20}. The model takes into account LL coupling and treats
only very weak dissipation. The main results of the present work are as
follows.

i) For moderately steep confining potential we presented a microscopic model
that improves the quasi-microscopic approach of Refs. \cite{11} and
\cite{18}. In particular, the model combines features of the following
distinct edge wave mechanisms. In mostly classical models, e.g., \cite{1},
\cite{7}, the position of the edge does not vary but the charge density
profile at the edge does. In a sence this edge wave mechanism is the analog
of that for the Kelvin wave \cite{21} at the edge of a rotating "shallow"
sea with chirality determined by the Coriolis parameter which corresponds to
the cyclotron frequency $\omega_{c}$. Another edge-wave mechanism,
fully quantum mechanical, is that of Refs. \cite{10}, \cite{17}, and
\cite{22}-\cite{23}, in which, for $\nu=1$, only the edge position of an
incompressible 2DEG of the lowest LL varies and with respect to which the
density profile is that of the undisturbed 2DEG. The approach of these
works is limited to the subspace of the lowest LL wave functions,
neglects LL mixing and dissipation, and results in a single EMP with
dispersion law similar to that in Ref. \cite{1}.

ii) We confirmed, for $\nu=1 (2)$, the existence of EMP modes in addition to
the fundamental EMP and single one obtained for vertically steep unperturbed
electron density profile \cite{1}. This is in line with our earlier
quasi-microscopic results \cite{11} for moderately steep \cite{11} confining
potentials. The additional modes result from an {\bf exact} solution of
Eq. (\ref{16}) in terms of the complete set of the Hermite polynomials.
With this expansion we can also make contact, for $\nu=1 (2)$, with other
microscopic theories \cite{10}, \cite{17}, and \cite{22}-\cite{23} that are
limited to the subspace of the lowest LL wave functions. If we retain only
the $n=0$ term in Eq. (\ref{16}), we obtain only the $n=0$ LL fundamental
mode with the same dispersion relation $\omega\sim q_{x}\ln q_{x}$. The
additional modes result from retaining ( correspond to) higher-order terms
in the expansion Eq. (\ref{16}). This shows that the limitation to this
LL subspace is too restrictive and indicates the importance of LL
coupling even for $\nu=1 (2)$.

iii) An important additional mode is the {\it dipole} EMP, presented in Sec.
III B, with dispersion relation $\omega\sim q_{x}^{3} $. This differs
markedly from the corresponding result $\omega\sim q_{x}$ of Ref. \cite{11}
and signals {\it nonlocal} responses that were previously neglected. This
and other differences between the present results and those of Ref. \cite{11}
further indicate the importance of having a fully {\it microscopic} theory
for EMPs.

iv) As we showed, taking LL coupling into account is an essential ingredient
especially for $\nu>2$. As Fig. 1 and Fig. 3 demonstrate, the coupling
strongly renormalizes the {\it uncoupled} fundamental LL modes and results
in one mode behaving as the usual one $\omega\sim q_{x}\ln q_{x}$ and the
others behaving in an {\it acoustic} manner $\omega\sim q_{x}$. This partly
reminds the results of Ref. \cite{7} but the resemblance should not be
overestimated because the two models are drastically different.

v) We have not treated spin effects beyond the HA or RPA. Our study is
focused on important wave effects of non-spin nature and spin splitting is
assumed negligible for even $\nu$. Thus, skyrmions and spin textures are not
dealt with in this work. Though neglecting the spin splitting is a
reasonable approximation, for even $\nu \geq 2$
%%% \geq 2 here better then > 2
in the bulk of the channel, its validity near the edges remains uncertain
in view of the work of Refs. \cite{16} and \cite{24}.

   (vi) Finally, a  few remarks are in order about  the studies
of Refs.  \cite{25} and \cite{26}  that used the RPA. 
The study of Ref. \cite{26} is too simplified and in essence  
repeats the results of Ref. \cite{25} and of works cited therein
for the "optical" EMP. As for the results of Ref. \cite{25}, 
due to omitted important contributions,
they are essentially different from ours for both the acoustic
and the optical modes. As Sec. IV A shows, an important logarithmic term,
caused by the Coulomb interaction, is missed
in Eqs.  (54a) and (54b) of Ref. \cite{25}. 
In addition,  Refs. \cite{25} and \cite{26} did not study a
dipole mode or   multipole modes pertaining to any occupied LL.
These modes of ours are totally different than those of
Secs. 3.1.6 and  3.1.7 of Ref. \cite{25}.

\acknowledgements
This work was supported by NSERC Grant No. OGP0121756. In addition, O G B
acknowledges partial support by the Ukrainian SFFI Grant No. 2.4/665.
%\clearpage
\ \newline
\appendix{APPENDIX}

The expressions for the coefficients $a_{k}(q_{x}),\ k=1,2,3 $ in Eq. (\ref
{53}) are as follows:

\begin{equation}
a_{2}(q_{x})=-[\tilde{a}_{00}(q_{x})+\tilde{a}_{11}^{11}(q_{x},0)+
\tilde{a}_{22}^{22}(q_{x},0)] ,  \label{54}
\end{equation}

\begin{eqnarray}
a_{1}(q_{x})=&& \tilde{a}_{00}(q_{x})\tilde{a}_{11}^{11}(q_{x},0)+ \tilde{a}%
_{00}(q_{x}) \tilde{a}_{22}^{22}(q_{x},0)+ \tilde{a}_{11}^{11}(q_{x},0)
\tilde{a}_{22}^{22}(q_{x},0)-  \nonumber \\
* \   \nonumber \\
&&[a_{00}^{22}(q_{x},\Delta y_{02})]^{2}- [a_{00}^{11}(q_{x},\Delta
y_{01})]^{2}- [a_{11}^{22}(q_{x},\Delta y_{12})]^{2} ,  \label{55}
\end{eqnarray}
and

\begin{eqnarray}
a_{0}(q_{x})=&& \tilde{a}_{00}(q_{x}) [a_{11}^{22}(q_{x},\Delta
y_{12})]^{2}+ \tilde{a}_{11}^{11}(q_{x},0) [a_{00}^{22}(q_{x},\Delta
y_{02})]^{2}+\tilde{a}_{22}^{22}(q_{x},0) [a_{00}^{11}(q_{x},\Delta
y_{01})]^{2}-  \nonumber \\
* \   \nonumber \\
&&\tilde{a}_{00}(q_{x}) \tilde{a}_{11}^{11}(q_{x},0) \tilde{a}%
_{22}^{22}(q_{x},0)-2 a_{00}^{11}(q_{x},\Delta y_{01})
a_{11}^{22}(q_{x},\Delta y_{12}) a_{00}^{22}(q_{x},\Delta y_{02}) .
\label{56}
\end{eqnarray}
Here $\tilde{a}_{00}(q_{x})=a_{00}(q_{x})+v_{g0}/ (2e^{2}/\pi \hbar \epsilon)
$, and $\tilde{a}_{ii}^{ii}(q_{x},0)=a_{ii}^{ii}(q_{x},0)+v_{gi}/
(2e^{2}/\pi \hbar \epsilon)$, $i=1,2$. We also have $a_{22}^{22}(q_{x},0)
\approx \ln(1/|q_{x}|\ell_{0})$ and, for $\Delta y_{02}, \Delta y_{12} \gg
\ell_{0}$, $a_{00}^{22}(q_{x},\Delta y_{02}) \approx \ln(1/|q_{x}|\ell_{0})-
\ln(\Delta y_{02}/\ell_{0})+0.1$, and $a_{11}^{22}(q_{x},\Delta y_{12})
\approx \ln(1/|q_{x}|\ell_{0})- \ln(\Delta y_{12}/\ell_{0})+0.1$.

Eq. (\ref{53}) corresponds to the irreducible case $Q(q_{x})=q^{3}+r^{2}<0$,
where $q=[a_{1}-a_{2}^{2}/3]/3$, $r=(a_{1}a_{2}-3a_{0})/6-a_{2}^{3}/27$.
With $\rho=\sqrt{r^{2}-Q}$ and $\theta=\arctan(\sqrt{-Q}/r)$ the roots obey
$\omega^{ `}_{1} +\omega^{ `}_{2}+\omega^{ `}_{3}=-a_{2}$ and are given by

\begin{equation}
\omega^{ `}_{1}=2 \rho^{1/3} \cos(\theta/3)-a_{2}/3 ,  \label{A1}
\end{equation}

\begin{equation}
\omega^{ `}_{2}=-\rho^{1/3} \cos(\theta/3)-a_{2}/3 -\sqrt{3} \rho^{1/3}
\sin(\theta/3) ,  \label{A2}
\end{equation}
and
\begin{equation}
\omega^{ `}_{3}=-\rho^{1/3} \cos(\theta/3)-a_{2}/3 +\sqrt{3} \rho^{1/3}
\sin(\theta/3) .  \label{A3}
\end{equation}

\begin{figure}
\caption{Dispersion relation for $\protect\nu=6$ and $B=3$Tesla. The dashed
curves are, from top to bottom, the {\bf decoupled} fundamental modes of the
$n=0$, $n=1$, and $n=2$ LLs, respectively. The solid and dotted curves are
for correspondingly the same {\bf coupled} modes, $\protect\omega_{*}=2e^{2}/%
\protect\pi \hbar \protect\epsilon \ell_{0}$. For the solid curves
approximation (Eq. ( \protect\ref{61})) is used.}
\label{fig.1}

\caption{Spatial structure of the renormalized fundamental modes
of the $n=0$ (solid curve),
$n=1$ (dotted curve), and $n=2$ (dashed curve) LLs, whose
dispersion laws are shown by the solid curves in Fig. 1.
The solid and dotted curves show
$\protect\rho(\protect\omega,q_{x},y)/\protect\rho_{*}$ and the
dashed curve
$\protect\rho(\protect\omega,q_{x},y)/2\protect\rho_{*}$ as a
function of $Y=(y-y_{r2})/\ell_0$ with $\protect\rho_{*}=%
\protect\rho_{0}^{(0)}(\protect\omega, q_x)/\sqrt{\protect\pi}\ell_0$.}
\label{fig.2}
%\end{figure}  %%% must be "\rho_{0}^{(0)}" not '\rho_{0}^{0}'

\caption{Dispersion relation for $\protect\nu=6$ and $B=1.5$Tesla.
All curves are marked as in Fig. 1.}
\label{fig.3}

\caption{Spatial structure,
$\protect\rho(\protect\omega,q_{x},y)/\protect\rho_{*}$, of the
renormalized fundamental modes of the $n=0$ (solid curve),
$n=1$ (dotted curve), and $n=2$ (dashed curve) LLs, whose dispersion laws
are shown by the solid curves in Fig. 3.}
\label{fig.4}
\end{figure}
\end{document}